\documentclass[10pt]{bmc_article}

\usepackage{cite} 
\usepackage{url}  
\usepackage{ifthen}  
\usepackage{multicol}   
\usepackage[utf8]{inputenc} 
\usepackage{longtable}
\usepackage{multirow}
\usepackage{slashbox}
\usepackage{graphicx}
\newboolean{publ}

\newenvironment{bmcformat}{\begin{raggedright}\baselineskip20pt\sloppy\setboolean{publ}{false}}{\end{raggedright}\baselineskip20pt\sloppy}

\begin{document}
\begin{bmcformat}


\title{Linear model for fast background subtraction in oligonucleotide microarrays}


      
\author{K. Myriam Kroll$^1$
\email{KMK: Myriam.Kroll@fys.kuleuven.be} 
Gerard T. Barkema$^{2,3}$ 
\email{GTB: G.T.Barkema@phys.uu.nl} 
and 
Enrico Carlon\correspondingauthor$^1$
\email{EC\correspondingauthor: Enrico.Carlon@fys.kuleuven.be} 
}


\address{%
    \iid(1)Institute for Theoretical Physics, Katholieke Universiteit
Leuven, Celestijnenlaan 200D, Leuven, Belgium\\
    \iid(2)Institute for Theoretical Physics, Utrecht University, Leuvenlaan 4, 
3584 CE Utrecht, The Netherlands\\
    \iid(3)Instituut-Lorentz for Theoretical Physics, University of Leiden,
Niels Bohrweg 2, 2333 CA Leiden, The Netherlands
}%

\maketitle


\begin{abstract}
        \paragraph*{Background:} One important preprocessing step in
        the analysis of microarray data is background subtraction. In
        high-density oligonucleotide arrays this is recognized as a
        crucial step for the global performance of the data analysis
        from raw intensities to expression values. 

        \paragraph*{Results:} We propose here an algorithm for background
        estimation based on a model in which the cost function is
        quadratic in a set of fitting parameters such that minimization
        can be performed through linear algebra. The model incorporates
        two effects: 1) Correlated intensities between neighboring
        features in the chip and 2) sequence-dependent affinities for
        non-specific hybridization fitted by an extended nearest-neighbor
        model.

        \paragraph*{Conclusions:} The algorithm has been tested on
        360 GeneChips from publicly available data of recent
        expression experiments. The algorithm is fast and accurate.
        Strong correlations between the fitted values for different
        experiments as well as between the free-energy parameters and
        their counterparts in aqueous solution indicate that the model
        captures a significant part of the underlying physical chemistry.
\end{abstract}

\ifthenelse{\boolean{publ}}{\begin{multicols}{2}}{}


\section*{Background}

The analysis of microarray data has attracted continuous interest over the
past years in the Bioinformatics community (see e.g. \cite{gent03}). The
problem consists in obtaining the gene expression level from the
experimental measurements, which are the emitted fluorescence intensities
from different sites in the array.  On general grounds one expects that
the experimental signal can be decomposed into two contributions:
\begin{equation} 
I = I_{\rm bg} + I_{\rm SP} (c)
\label{eq:01}
\end{equation}
where $I_{\rm SP} (c)$ is the specific signal due to the hybridization
of the surface-bound probe sequence with a complementary target
sequence. This quantity depends on the concentration $c$ of the
complementary strand in solution (target). The non-specific term $I_{\rm
bg}$ has different origins. It arises due to spurious effects such as 
incomplete hybridization where probe sequences bind to only partially 
complementary targets or due to other optical effects.

Models based upon the physical chemistry of hybridization (see
e.g. \cite{held03}) predict a linear increase of the specific signal
until saturation is approached. In case of highly expressed genes
the specific part of the signal $I_{\rm SP}(c)$ dominates the total
signal intensity $I$ and hence one can safely make the approximation
$I \approx I_{\rm SP} (c)$. For lowly expressed genes, as well as for
sequences with a low binding affinity, the specific and the non-specific
contribution to the total intensity can be of comparable magnitude. In
this case an accurate estimate of $I_{\rm bg}$ is crucial to draw
reliable conclusions concerning the expression level; estimates based
on the intensity distribution over the whole chip suggest that this is
the case for roughly a quarter or half of the probes~\cite{ferr09}. Once
the background is calculated the gene expression level is then computed
from background subtracted data $I - I_{\rm bg}$.

In this paper we present an algorithm for the calculation of the
background level for Affymetrix expression arrays, also known as
GeneChips.  In these arrays the probe sequences come in pairs: for
each perfect match (PM) probe, which is exactly complementary to the
transcript sequence in solution, there is a second probe with a single
non-complementary nucleotide with respect to the specific target. The
latter is called mismatch (MM) probe.

Several algorithms for background analysis of Affymetrix chips are
available. Some of these use the MM intensities as corrections for
non-specific hybridization, while others rely on PM intensities only. For
instance, the Affymetrix MAS 5.0 software (Microarray Analysis Suite
version 5.0) uses the difference $(I^{\rm PM} - I^{\rm MM})$ as estimator
of the specific signal; an adjusted MM intensity (ideal MM) is used in the
case the MM intensity exceeds the PM signal~\cite{affy01}. The
Robust Multiarray Algorithm (RMA)~\cite{iriz03} uses a different type
of subtraction scheme which does not involve the MM intensities. The
more recent version of this algorithm GCRMA performs background
subtraction using information on the probe sequence composition
through the calculation of binding affinities~\cite{wu04}. The position-dependent
nearest-neighbor model (PDNN)~\cite{zhan03} fits the background intensity
using weight factors which depend on the position along the probe. The free
energy parameters then enter in a nonlinear function. In the VSN
algorithm~\cite{hube02} a generalized log transform is used to background
correct the data. A study dedicated to the performance of different
algorithms showed that the type of background subtraction used has a large
effect on the global performance of the algorithms~\cite{iriz06}. It is
therefore not surprising that the background issue has attracted a lot
of interest by the scientific community.

In this paper we present an algorithm for background estimation
which combines information from the sequence composition and physical
neighbors on the chip. This algorithm relies on previous work by the
authors~\cite{krol08}. While the previous algorithm performed well with
respect to the accuracy of the background estimation, the computational
effort (per probe) involved was a severe limiting factor concerning its
practical usability.  The main cause of this significant computational
effort was the iterative minimization of a cost function with nonlinear
terms. The algorithm presented in this work involves a different cost
function which is quadratic in the parameters.  Its minimization can
be performed via standard matrix computations of linear algebra. The
algorithm is fast and accurate and is therefore suited for large
scale analysis.

This paper is organized as follows. In Methods we discuss the optimization
step from singular value decomposition and we provide the details of the
selected cost function.  In Results a test of the algorithm on about 360
Genechips from recent (2006 onwards) experiments from the Gene Expression
Omnibus (GEO) is presented. Finally, the advantages of this scheme and
its overall performance as background subtraction method is highlighted.

\section*{Methods}

\subsection*{Approach}

The general assumption is that the (natural) logarithm of the background
intensity can be approximated by a function linear in some fitting
parameters $\omega_{\alpha}$. Once these parameters are set to their
optimal values ${\bar \omega}_\alpha$ an estimate of the background
intensity for the $i$-th ($i=1,2,\ldots N_{dim}$) PM probe can be
obtained as
\begin{equation}\label{eq:back_est}
\log I^{\rm est}_{i, {\rm bg}} = 
\sum_{\alpha}  \Omega_{i \alpha} {\bar \omega}_\alpha,\quad\quad
\alpha=1,2,\ldots N_f
\end{equation}
where $N_f$ is the number of fitting parameters. $\Omega_{i \alpha}$
is a sequence- and position-dependent element of the $N_{dim}\times
N_f$-dimensional matrix $\Omega$, which will be defined below.

The optimized values for the model parameters $\bar{\omega}_\alpha$
are obtained by minimizing the difference (of the logarithms) of the
observed and estimated intensities
\begin{eqnarray}\label{eq:function_s}
S &=&  \frac{1}{K}\sum_{k\in\mathcal{K}} \left(\log I_k - \sum_\alpha \Omega_{k
\alpha} \omega_\alpha
\right)^2
\end{eqnarray}
i.e.~solving a linear least square problem.  The sum extends over the
training set $\mathcal{K}$ which is a subset (with $K$ elements) of
the intensities of all annotated features. The choice of the elements
of $\mathcal{K}$ will be discussed later (see Data Set - Parameter
Optimization).  The minimum is found by imposing vanishing partial
derivatives of $S$ w.r.t. $\omega_\alpha$. This yields the following $N_f$
linear equations

\begin{equation}
\label{eq:lin_eq}
\left. \frac{\partial S}{\partial \omega_\alpha} \right|_{\bar\omega} = 
\sum_{k\in\mathcal{K}} \Omega_{k \alpha} \left(\log  I_k-\sum_\beta
\Omega_{k\beta} 
\omega_\beta \right) = 0,
\end{equation}

which can be rewritten as
\begin{equation}\label{eq:normaleq}
\sum_\beta  M_{\alpha \beta} \bar \omega_\beta = \Gamma_\alpha. 
\end{equation}
Here, we have introduced the matrix
\begin{equation}\label{eq:matrixM}
M_{\alpha \beta} = \sum_{k\in\mathcal{K}} \Omega_{k \alpha} \Omega_{k \beta}
\end{equation}
and the vector $\Gamma_\alpha$
\begin{equation}\label{eq:vectorGamma}
\Gamma_{\alpha} = \sum_{k\in\mathcal{K}} \Omega_{k \alpha} \log I_k.
\end{equation}
(Note that $M = \Omega^{T} \Omega$ is symmetric and $\dim(M)=N_{f}\times
N_{f}$).

The solution of Eq.~(\ref{eq:normaleq}) is given by the vector
\begin{equation}\label{eq:sol_omegabar}
\bar\omega = M^{-1} \Gamma
\end{equation}
of the optimal parameter values. If the matrix $M$ is singular,
$M^{-1}$ has to be replaced by its pseudoinverse $M^{+}$ which can
be obtained by means of Singular value decomposition (SVD).  In this
work a standard SVD algorithm based on Golub and Reinsch is used (see
e.g. \cite{golu70,golu96}).

Due to the symmetry of $M$ only half of the off-diagonal elements need
to be generated, hence reducing the computational effort. For the chips
tested in this paper with dimensions up to $1164 \times 1164$ features,
the computational time on a standard PC (x86\_64 Intel Core 2 Duo with
3GHz, 3 GB RAM) required to estimate the background intensities is 8
to 10 seconds for the larger chips, and faster for the smaller ones.
This makes our algorithm an order of magnitude faster than our previous
version \cite{krol08}, 3 to 5 times faster than GC-RMA, PDNN and MAS5,
and about twice as slow as RMA and DFCM (Bioconductor packages were
used for the testing).  Note that for our algorithm, the time estimate
includes both reading in the CEL-file and the background calculation,
as it is done in one step.

This computation involves the generation of the matrix $M$ and vector
$\Gamma$ (from Eqs.~(\ref{eq:matrixM}) and (\ref{eq:vectorGamma})), the
SVD of $M$ to solve Eq.~(\ref{eq:normaleq}) and the estimation of the
background intensity for all PM probes through Eq.~(\ref{eq:back_est}).
Differently from other approaches in which the cost function is
minimized by means of Monte Carlo methods~\cite{ono08} or other dynamical
algorithms~\cite{krol08}, the solution of SVD provides the exact minimum
of the cost function Eq.~(\ref{eq:function_s}). Hence, there is no risk
in getting stuck in local minima different from the global one. 

\subsection*{Data Set - Parameter Optimization}

As mentioned above, probes in Affymetrix form PM/MM pairs. Consider
now a target sequence at a concentration $c$ in solution. The analysis
of Affymetrix spike-in data (see e.g. \cite{burd06}) shows that not only the
PM signal increases with increased target concentration $c$ but also
the MM intensity. This is an indication that a single MM nucleotide
only partially prevents probe-target hybridization. Therefore the
intensity of MM probes can also be decomposed in a non-specific
and specific part as in Eq.~(\ref{eq:01}). Supported by Affymetrix
spike-in data analysis, our assumption is that the non-specific part of
the hybridization is about equal for PM and MM probes: $I_{bg}^{(PM)}
\approx I_{bg}^{(MM)}$. The specific part of the signal is different in
those two cases; equilibrium thermodynamics suggests a constant ratio
$I_{SP}^{(PM)}(c)/I_{SP}^{(MM)}(c) = const. > 1$, independent of the
target concentration, as observed in experiments~\cite{ferr09}.

These insights are useful for the selection of probes for the optimization
set $\mathcal{K}$ in Eq.~(\ref{eq:function_s}): $\mathcal{K}$ includes
all MM probes whose intensities are below a certain threshold $I_0$
and whose corresponding PM intensities also fulfill $I^{\rm PM}_k <
I_{0}$ (a similar selection criterion was recently used by Chen et
al.~\cite{chen09}). The threshold $I_0$ is chosen on the basis of
the total distribution of the intensities. $\mathcal{K}$ contains a
significant fraction of the mismatch probes: typically $35\%$.
Since the specific signal of MM intensities is lower than that of
their corresponding PM's, they provide more reliable information on the
background. The coordinates and sequences of the probes in $\mathcal{K}$
are then fitted to the intensities of these probes yielding the parameters
$\omega$. With those newly acquired parameters $\omega$ the background
signal of all MM probes is estimated base upon the assumption $I_{bg}^{(PM)}
\approx I_{bg}^{(MM)}$.

\subsection*{The matrix $\Omega$}

The choice of the matrix elements of $\Omega$ is dictated by input from
physical chemistry as well as by the architecture of the microarray.
Different schemes involving different choices for $\Omega$ with
a varying number of parameters $N_f$ were tested.  Given a choice of
$\Omega$ and in particular the number $N_f$ of fitting parameters, the
accuracy of the background estimation is reflected by the value of $S$
from the minimization of Eq.~(\ref{eq:function_s}). While the addition
of fitting parameters always yields lower values of $S$, a too large set
of fitting parameters runs the risk of ``overfitting". The final choice
of $\Omega$ is a compromise between a minimization of $S$ and the use of the
smallest possible set of parameters.

In the present model the number of parameters is $N_f=50$.
Similarly to the previous work \cite{krol08} these parameters can be
split into two groups: a first group describes the correlation of the
background intensities with features which are physical neighbors on the
chip; the second group are nearest-neighbor parameters which describe
affinities for non-specific hybridization to the chip.

\subsubsection*{Physical Neighbors on the Chip}

The first $18$ parameters $\omega_\alpha$ describe the correlation of the
background intensity with physical neighboring sequences in the array.
Let $(x_i,y_i)$ be the coordinate of the $i$-th MM sequence $s(i)$. Then,
its eight neighbors are located at $(x_i\pm 1,y_i)$, $(x_i \pm 1,y_i\pm1)$,
and $(x_i, y_i\pm 1)$.
As there is evidence that base-specific interactions (purine/pyrimidine
asymmetry) might influence the hybridization process in
general~\cite{bind05}, we furthermore distinguish between probes
whose central nucleotide is either A/G (purine) or C/T (pyrimidine).
Then, the corresponding matrix elements in case of purines can be
written as
\begin{eqnarray}
\Omega_{i 1} &=& \delta_{s(i)}^{A/G}
\label{eq:alpha1} \\
\Omega_{i \alpha} &=& \delta_{s(i)}^{A/G} \log I (x_i+p_\alpha,
y_i+q_\alpha),
\quad \ \ 2 \leq \alpha \leq 9
\label{eq:alpha2_9}
\end{eqnarray}
with
\begin{equation}\nonumber
  \delta_{s(i)}^{A/G} =
\left\{
\begin{array}{cl}
1 &\rm{if\ the\ 13^{th}\ nucleotide\ is\ purine\ (A\ or\ G)} \\
0 &\rm{otherwise}
\end{array}
\right.
\end{equation}
and
\begin{eqnarray}\nonumber
 p_2=p_4=p_7=q_7=q_8=q_9&=&1\\ \nonumber
 p_6=p_9=q_2=q_3&=&0\\\nonumber
 p_3=p_5=p_8=q_4=q_5=q_6&=&-1,\nonumber
\end{eqnarray}
so that the intensities of the neighboring features explicitly enter
the calculation of the background intensity of $(x_i,y_i)$ as matrix
elements $\Omega_{i \alpha}$ ($2 \leq \alpha \leq 9$).  In analogy to
Eqs.~(\ref{eq:alpha1},\ref{eq:alpha2_9}) we define $\Omega_{i\alpha}$
with $10\leq\alpha\leq 18$ corresponding to the sequences with a central
pyrimidine.

\subsubsection*{Nearest-Neighbor Free Energy Parameters}

The second contribution to the background model arises from the sequence
composition. Let us first label the 16 dinucleotides according to the
order $\{CC, GC, AC, TC, CG, GG, AG\ldots TT\}$. We then define
the next 16 matrix elements as
\begin{eqnarray}
\Omega_{i \alpha} &=& \sum_{l=1}^{24} \delta_{l, l+1}^{\alpha-18} (s(i)),
\quad \quad \quad\quad
19 \leq \alpha \leq 34
\label{eq:alpha19_35}
\end{eqnarray}
where
\begin{equation}\nonumber
\delta_{l, l+1}^\gamma (s) =
\left\{
\begin{array}{cl}
1 &\rm{if\ nucleotides\ at\ l\ and\ l+1\ form\ pair\ of\ type}\ \gamma \\
0 &\rm{otherwise}
\end{array}
\right.
\end{equation}
according to the order given above. The sum runs over all the 24
dinucleotides along a probe sequence. The matrix element
$\Omega_{\alpha i}$ is equal to the number of dinucleotides of a given
type in the sequence $s(i)$. For instance, if the sequence $s(i)$
contains $4$ dinucleotides of type $CC$ and $2$ of type $GC$, then
$\Omega_{i, 19} = 4$ and $\Omega_{i, 20} = 2$. Hybridization
thermodynamics predicts $\log I \propto \Delta G$ where $\Delta
G$ is the hybridization free energy. In the nearest-neighbor model
\cite{bloo00} the free energy is written as a sum of dinucleotide terms.
Therefore, the parameters $\omega_\alpha$ ($19 \leq \alpha \leq 35$) are
the analogues of the free energy parameters of the nearest-neighbor model.

\subsubsection*{Position-Dependent Nearest-Neighbors}

Finally, we consider the possibility that the hybridization strength is
``modulated" along the sequence by a parabolic weight as
\begin{eqnarray}
\Omega_{i \alpha} &=& \sum_{l=1}^{24} \delta_{l, l+1}^{\alpha-35} (s(i))
\ (l - l_m)^2,
\quad\quad
35 \leq \alpha \leq 50
\label{eq:alpha35_50}
\end{eqnarray}
where $l_m=12.5$,~i.e. each dinucleotide is given a parabolic weight
according to its position relative to the center at $l_m$ of the
sequence. Thus, possible ``unzipping" effects of the DNA-RNA duplex
are approximately accounted for by Eq.~(\ref{eq:alpha35_50}).

The introduction of a position-dependence effect is in analogy with work
done by other groups~\cite{zhan03,naef03,bind05,zhan07}. However, we
do not introduce a position-dependent weight for each position along the
25-mer sequences. Instead, we limit ourselves to a parabolic modulation
of the parameters along the chain, which drastically reduces the number
of parameters involved in the model.

\subsubsection*{Invariances}

Given the definition of $\Omega$ above it can be shown that
Eq.~(\ref{eq:normaleq}) permits multiple solutions. Therefore, the optimal
parameters $\bar\omega_\alpha$ are not unique.  Consider a set of optimal
parameters $\bar{\omega}_\alpha$ which minimizes the cost function $S$
(Eq.~(\ref{eq:function_s})). Let us then add a constant $\lambda$
to the $16$ nearest-neighbor parameters
\begin{equation}
\bar{\omega}'_\alpha = \bar{\omega}_\alpha + \lambda.
\ \ \ \ \ \ \ \ \ \ \ \ \ \ \ \ 
(19 \leq \alpha \leq 34)
\label{eq:inv1}
\end{equation}
From Eq.~(\ref{eq:alpha19_35}) we obtain
\begin{equation}
\sum_{\alpha =19}^{34} \Omega_{i \alpha} \bar{\omega}'_\alpha = 
\sum_{\alpha =19}^{34} \Omega_{i \alpha} \bar{\omega}_\alpha + 24 \lambda
\label{eq:shift1}
\end{equation}
because $\Omega_{i \alpha}$ counts the frequency of the dinucleotide
$\alpha$ in the sequence corresponding to feature $i$ and because there are $24$
dinucleotides in a $25$-mer probe sequence. Now, also consider
\begin{equation}
\bar{\omega}'_1 = \bar{\omega}_1 - 24 \lambda
\ \ \ \ \ \ \ \ \ \ \ \ \ \ \ \ 
\bar{\omega}'_{10} = \bar{\omega}_{10} - 24 \lambda,
\label{eq:shift2}
\end{equation}
while $\bar{\omega}'_\alpha = \bar{\omega}_\alpha$ for all other
$\alpha$.  We conclude that the reparametrization of~Eqs.~(\ref{eq:inv1})
and~(\ref{eq:shift2}) yields
\begin{equation}
\sum_{\alpha =1}^{50} \Omega_{i \alpha} \bar{\omega}'_\alpha = 
\sum_{\alpha =1}^{50} \Omega_{i \alpha} \bar{\omega}_\alpha 
\label{eq:inv2}
\end{equation}
since the shifting of Eq.~(\ref{eq:shift2}) compensates the one introduced
by Eq.~(\ref{eq:shift1}). This reparametrization, valid for any real
$\lambda$, leaves $S$ invariant, and produces a zero eigenvalue of the
matrix $M$ of Eq.~(\ref{eq:matrixM}).

Similarly, one can verify that there is at least a second zero
eigenvalue: a shift of the position-dependent nearest-neighbor
parameters $\bar{\omega}_\alpha' = \bar{\omega}_\alpha + \lambda$
(for $35 \leq \alpha \leq 50$) as well as of $\bar{\omega}_1' =
\bar{\omega}_1 - 1150\lambda$, $\bar{\omega}_{10}' = \bar{\omega}_{10}
- 1150\lambda$ leaves $S$ invariant. To obtain the latter equations
Eq.~(\ref{eq:alpha35_50}) and $\sum_{l=1}^{24}(l-l_m)^2=1150$ have to
be applied.

Having zero eigenvalues, the matrix $M$ is therefore not invertible;
the SVD thus provides the appropriate pseudo-inverse as discussed above.
Accidental degeneracies or quasi-degeneracies of M could also occur,
yielding eigenvalues close to zero in machine precision. These are, however,
rare, and were actually never found in the calculations presented here.

\section*{Results} 

We analyzed a total of 366 CEL-files which are publicly available from the
GEO server (www.ncbi.nlm.nih.gov/geo). Table~\ref{Tab:overview_organisms_en} gives an overview of the
distribution of CEL-files over the twelve different organisms considered
in this study. The array size for an organism might vary depending on
the GSE accession number, since the most recent Affymetrix chips tend
to use smaller features, thus more probes can be accommodated on the
same surface area. For instance the Human HGU-133A contains $712^2$
features, while the Human Genome U133 Plus 2.0 Array goes to $1164^2$
features. The last column of Table~\ref{Tab:overview_organisms_en} lists the attained minimum value
$\bar S_{min}$, which ranges from $0.017$ for Escheria Coli to $0.11$
for Oryza Sativa.  $\bar S_{min}$ estimates the mean squared deviation
of the logarithm of the estimated background intensity from the actual
background value. For instance $\bar S_{min}=0.01$ corresponds to a
$10\%$ deviation, while $\bar S_{min}=0.1$ corresponds to a $37\%$
deviation.  

\begin{table}[!t]
\centering
{\begin{tabular}{lllcl}
 Organism & GEO \# & Chiptype (dimension) & \# files & $\bar S_{min}$ \\
\hline
&GSE4847         & ATH1-121501 (712x712)&     18      &       0.0259\\
{A. Thaliana} 
&GSE7642        &  ATH1-121501 (712x712) &   12      &       0.0544\\
&GSE9311        &   ATH1-121501 (712x712) &  8       &       0.0546\\
\hline
{C. Elegans} &
GSE6547         &   Celegans (712x712)  & 25      &       0.0361\\
&GSE8159        &   Celegans (712x712)   &7       &       0.0396\\
\hline
{D. Melanogaster} &
GSE3990         &   Drosophila\_2 (732x732)    & 6       &       0.0620\\
&GSE6558        &    DrosGenome1  (640x640) &24      &       0.0605\\
\hline
D. Rerio &GSE4859      & Zebrafish (712x712) &       8       &       0.0357\\
\hline
&GSE11779&  E\_coli\_2 (478x478)  &  3       &       0.0869\\
{E. Coli} 
&GSE2928        &   Ecoli (544x544)   &12      &       0.0172\\
&GSE6195        &    E\_coli\_2 (478x478)  &4       &       0.0664\\
\hline
&GSE10433&  HG-U133A\_2 (732x732)  &  12      &       0.0757\\
&GSE5054        &    HG-U133A (712x712)&20      &       0.0392\\
{H. Sapiens}
&  & HG-U133A\_2 (732x732) &  & \\
&GSE7148        &    HG-U133A (712x712)  &14      &       0.0296\\
&GSE8514        &    HG-U133\_Plus\_2 (1164x1164)  &15      &       0.0738\\
\hline
&GSE11897&   MOE430A (712x712)& 11      &       0.0640\\
&  & MOE430B (712x712) &  & \\
{M. Musculus}
&  & Mouse430\_2 (1002x1002) &  & \\
&GSE6210        &   Mouse430\_2    &12      &       0.0594\\
&GSE6297        &   Mouse430\_2    &24      &       0.0325\\
\hline
O. Sativa & GSE15071 & Rice (1164x1164)  &20      &       0.1157\\
\hline
&GSE4494 &   RG\_U34A (534x534)   &59      &       0.0488\\
{R. Norvegicus} 
&GSE7493        &   Rat230\_2 (834x834)  & 9       &       0.0497\\
&GSE8238        &   Rat230\_2 (834x834)  & 4       &       0.0640\\
\hline
S. Aureus & GSE7944   & S\_aureus (602x602) &       6       &       0.0746\\
\hline
{S. Cerevisiae} &
GSE6073 &    YG\_S98 (534x534) & 12      &       0.0283\\
&GSE8379        &   YG\_S98 (534x534)  & 8       &       0.0180\\
\hline
X. Laevis & GSE3368   & Xenopus\_laevis (712x712) &       20      &       0.0514\\
\hline  
\end{tabular}}
\caption{Overview over organisms and number of CEL-files
 analyzed}
\label{Tab:overview_organisms_en}
\end{table}

Table~\ref{Tab:params1} provides a summary of the optimal parameters as
calculated from the Singular Value Decomposition for the minimization
of the cost function of Eq.~(\ref{eq:function_s}).  The parameters
$\bar{\omega}_1$ and $\bar{\omega}_{10}$ are associated to constant
intensities for purines and pyrimidines. Their magnitudes are not unique
due to the reparametrization as discussed in Eqs.~(\ref{eq:inv1}) to
(\ref{eq:inv2}). If information on neighboring probes is disregarded,
the value of $S$ typically increases by 45\%; if the sequence information
is not directly used, then it will increase by 52\%. In the following we
analyze the parameters associated to the local physical neighbors
and to the nearest-neighbor free energy.

\begin{table}
\centering
\scriptsize
\begin{tabular}{c|ccc|cc|cc|c}
&\multicolumn{3}{c}{A. Thaliana} &\multicolumn{2}{c}{C. Elegans}
&\multicolumn{2}{c}{D. Melanogaster} &\multicolumn{1}{c}{D. Rerio} 
\\
GEO no  & GSE4847 & GSE7642 & GSE9311 & GSE6547 & GSE8159 & GSE3990 & GSE6558 &
GSE4859 \\
\hline
$\bar{\omega}_1$        &0.178  &0.196  &0.261  &0.379  &0.520  &0.270  &0.421  &0.338
\\
$\bar{\omega}_2$        &0.050  &0.056  &0.042  &0.047  &0.050  &0.022  &0.025  &0.041
\\
$\bar{\omega}_3$        &0.051  &0.056  &0.041  &0.034  &0.042  &0.024  &0.024  &0.050
\\
$\bar{\omega}_4$        &-0.013 &-0.012 &-0.013 &-0.012 &-0.016 &-0.005 &-0.011 &-0.012
\\
$\bar{\omega}_5$        &-0.013 &-0.011 &-0.012 &-0.010 &-0.012 &-0.003 &-0.010 &-0.008
\\
$\bar{\omega}_6$        &0.186  &0.198  &0.224  &0.168  &0.271  &0.228  &0.195  &0.140
\\
$\bar{\omega}_7$        &0.004  &0.005  &0.003  &0.003  &0.005  &0.001  &0.005  &0.006
\\
$\bar{\omega}_8$        &0.002  &0.009  &0.003  &0.002  &0.005  &0.001  &0.006  &0.002
\\
$\bar{\omega}_9$        &0.013  &0.027  &0.010  &0.008  &0.008  &0.009  &0.012  &0.014
\\
$\bar{\omega}_{10}$     &-0.174 &-0.192 &-0.258 &-0.375 &-0.517 &-0.267 &-0.418 &-0.334
\\
$\bar{\omega}_{11}$     &0.063  &0.060  &0.059  &0.058  &0.062  &0.042  &0.057  &0.068
\\
$\bar{\omega}_{12}$     &0.069  &0.064  &0.064  &0.052  &0.059  &0.046  &0.061  &0.075
\\
$\bar{\omega}_{13}$     &-0.017 &-0.015 &-0.016 &-0.011 &-0.012 &-0.012 &-0.013 &-0.016
\\
$\bar{\omega}_{14}$     &-0.018 &-0.016 &-0.019 &-0.010 &-0.009 &-0.011 &-0.014 &-0.012
\\
$\bar{\omega}_{15}$     &0.258  &0.299  &0.328  &0.301  &0.459  &0.336  &0.316  &0.246
\\
$\bar{\omega}_{16}$     &0.004  &0.006  &0.005  &0.005  &0.009  &0.002  &0.008  &0.007
\\
$\bar{\omega}_{17}$     &0.002  &0.007  &0.003  &0.003  &0.006  &0.001  &0.007  &0.004
\\
$\bar{\omega}_{18}$     &0.013  &0.025  &0.012  &0.012  &0.015  &0.009  &0.016  &0.016
\\
\hline
$\bar{\omega}_{19}$ & 0.068 & 0.138 & 0.143 & 0.149 & 0.155 & 0.159 & 0.217 & 0.140
\\
$\bar{\omega}_{20}$ & 0.052 & 0.191 & 0.169 & 0.090 & 0.157 & 0.124 & 0.152 & 0.140 
\\
$\bar{\omega}_{21}$ & -0.021 & -0.038 & 0.006 & -0.033 & 0.030 & -0.060 & -0.033 & -0.069 
\\
$\bar{\omega}_{22}$ & -0.021 & 0.051 & 0.056 & -0.025 & 0.072 & -0.084 & 0.003 & -0.068 
\\
$\bar{\omega}_{23}$ & -0.032 & -0.162 & -0.147 & -0.038 & -0.121 & -0.060 & -0.094 & -0.104 
\\
$\bar{\omega}_{24}$  & 0.041 & 0.075 & 0.072 & 0.024 & 0.015 & 0.030 & 0.048 & 0.035 
\\
$\bar{\omega}_{25}$ & -0.063 & -0.221 & -0.162 & -0.122 & -0.140 & -0.172 & -0.200 & -0.190 
\\
$\bar{\omega}_{26}$ & -0.060 & -0.130 & -0.119 & -0.118 & -0.109 & -0.173 & -0.154 & -0.181 
\\
$\bar{\omega}_{27}$ & 0.000 & -0.008 & -0.047 & 0.035 & -0.041 & 0.052 & 0.035 & 0.055
\\
$\bar{\omega}_{28}$ & 0.036 & 0.148 & 0.076 & 0.062 & 0.063 & 0.105 & 0.096 & 0.128 
\\
$\bar{\omega}_{29}$ & -0.026 & -0.058 & -0.069 & -0.046 & -0.044 & -0.065 & -0.085 & -0.052 
\\
$\bar{\omega}_{30}$ & -0.050 & -0.032 & -0.093 & -0.065 & -0.052 & -0.106 & -0.102 & -0.078 
\\
$\bar{\omega}_{31}$ & 0.057 & 0.037 & 0.056 & 0.082 & 0.012 & 0.133 & 0.103 & 0.130 
\\
$\bar{\omega}_{32}$ & 0.093 & 0.180 & 0.173 & 0.125 & 0.112 & 0.211 & 0.181 & 0.221 
\\
$\bar{\omega}_{33}$ & 0.011 & -0.074 & -0.021 & -0.008 & -0.026 & 0.015 & -0.036 & 0.001 
\\
$\bar{\omega}_{34}$ & -0.009 & -0.017 & -0.016 & -0.023 & -0.014 & -0.035 & -0.047 & -0.022 
\\
Corr. Coeff   & 0.775 & 0.686 & 0.738 & 0.814 & 0.700 & 0.754 & 0.826 & 0.705 
\\

\hline
$\bar{\omega}_{35}$ & 0.020 & 0.016 & 0.016 & 0.019 & 0.013 & 0.016 & 0.012 & 0.018 
\\
$\bar{\omega}_{36}$ & 0.020 & 0.010 & 0.012 & 0.021 & 0.011 & 0.016 & 0.016 & 0.017 
\\
$\bar{\omega}_{37}$ & 0.025 & 0.025 & 0.022 & 0.029 & 0.019 & 0.028 & 0.027 & 0.031 
\\
$\bar{\omega}_{38}$ & 0.024 & 0.019 & 0.019 & 0.028 & 0.016 & 0.028 & 0.023 & 0.031 
\\
$\bar{\omega}_{39}$ & 0.026 & 0.036 & 0.034 & 0.030 & 0.030 & 0.027 & 0.031 & 0.033 
\\
$\bar{\omega}_{40}$  & 0.021 & 0.021 & 0.020 & 0.026 & 0.021 & 0.022 & 0.023 & 0.026 
\\
$\bar{\omega}_{41}$ & 0.028 & 0.039 & 0.035 & 0.036 & 0.031 & 0.035 & 0.038 & 0.039 
\\
$\bar{\omega}_{42}$ & 0.027 & 0.032 & 0.031 & 0.035 & 0.028 & 0.033 & 0.033 & 0.038 
\\
$\bar{\omega}_{43}$ & 0.023 & 0.024 & 0.026 & 0.024 & 0.023 & 0.019 & 0.021 & 0.021 
\\
$\bar{\omega}_{44}$ & 0.020 & 0.013 & 0.017 & 0.022 & 0.015 & 0.015 & 0.017 & 0.016 
\\
$\bar{\omega}_{45}$ & 0.024 & 0.028 & 0.028 & 0.030 & 0.024 & 0.027 & 0.029 & 0.029 
\\
$\bar{\omega}_{46}$ & 0.025 & 0.025 & 0.029 & 0.030 & 0.024 & 0.028 & 0.029 & 0.030 
\\
$\bar{\omega}_{47}$ & 0.020 & 0.022 & 0.020 & 0.022 & 0.021 & 0.015 & 0.019 & 0.016 
\\
$\bar{\omega}_{48}$ & 0.017 & 0.012 & 0.012 & 0.019 & 0.013 & 0.010 & 0.014 & 0.011 
\\
$\bar{\omega}_{49}$ & 0.023 & 0.029 & 0.026 & 0.028 & 0.023 & 0.023 & 0.028 & 0.026 
\\
$\bar{\omega}_{50}$ & 0.023 & 0.023 & 0.023 & 0.028 & 0.021 & 0.024 & 0.026 & 0.026 
\\
Corr coeff & -0.679 & -0.618 & -0.679 & -0.759 & -0.631 & -0.683 & -0.783 &
-0.658 
\\
\end{tabular}{}
\caption{Optimized parameter values as obtained from the minimization
of Eq.~(\ref{eq:function_s}). Magnitudes and signs of the parameters are
approximately constant across different experiments. The correlation
coefficients are the Pearson correlations between the nearest-neighbor
parameters and the free energy parameters measured in hybridization
experiments in aqueous solution \cite{sugi95}.}
\label{Tab:params1}
\end{table}

\subsection*{Parameters of physical neighbors in the Chip}

The parameters $\bar{\omega}_{2}$ to $\bar{\omega}_{9}$ and
$\bar{\omega}_{11}$ to $\bar{\omega}_{18}$ describe the coupling of
the background intensities to the physically neighboring features on the
chip. As already mentioned, our estimate of the PM background is based
on the non-specific intensity of the MM sequence. An Affymetrix chip
is designed such that MM and PM are found in rows at equal $y$-coordinates.
In addition, given a PM at $(x, y)$, the corresponding MM feature is at
$(x, y+1)$.

Figure~\ref{fig:neighbors_350} schematically represents the influence
of the neighboring intensities on the background value. The strength
of the correlation of the eight neighboring spots (i.e.~the magnitude
and sign of the corresponding $\bar \omega_\alpha$) with the central MM
feature is indicated by the color.
\begin{figure}[h!]
\centerline{\includegraphics{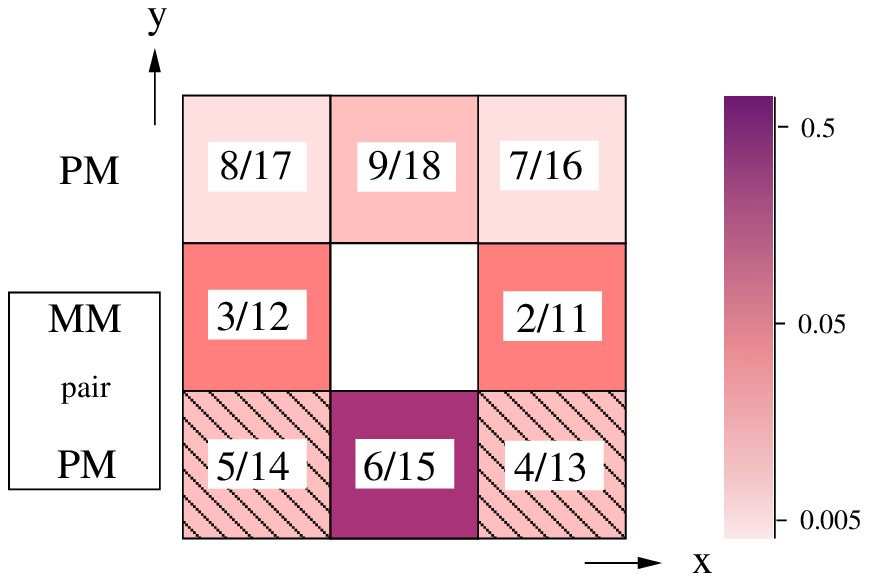}}
\caption{\textbf{Correlations with neighboring features:}
Schematic representation of the neighboring parameters on the
array. The inset numbers indicate the corresponding parameter(s)
$\bar{\omega}_{\alpha}$,~e.g. 3/12 represents $\bar{\omega}_{3}$,
$\bar{\omega}_{12}$ respectively. $\bar{\omega}_2$ through
$\bar{\omega}_{9}$ are related to purines, $\bar{\omega}_{11}$
to $\bar{\omega}_{18}$ to pyrimidines. The bar to the right gives the
intensity scale of the parameters.  The dashed pattern indicates
negative values. Box on the left indicates the rows of corresponding
PM and MM pairs. The central feature (position $(x,y)$) is a MM;
its corresponding PM is located just below it $(x,y-1)$, its intensity
has the strongest correlation with the central MM feature. The features
in $(x \pm 1,y-1)$}
\label{fig:neighbors_350}
\end{figure}


The numbers in Fig.~\ref{fig:neighbors_350} identify the associated parameters $\bar{\omega}_\alpha$.
For example $\bar{\omega}_6$ and $\bar{\omega}_{15}$ are associated to the intensity at $(x,y-1)$
with respect to the reference MM intensity with coordinates $(x,y)$.
There does not seem to be any evidence that the middle-nucleotide
classification in purines and pyrimidines reveals any insight concerning
the background. Instead, our results show that the absolute values of two
''corresponding`` (purine-pyrimidine pair) $\bar{\omega}_{\alpha}$'s
are generally of the same order of magnitude.  Across all species,
we find typical average outputs of
\begin{eqnarray}\nonumber
&\lbrace\bar{\omega}_{2/11},\ldots,\bar{\omega}_{9/18}\rbrace\approx
&\lbrace 0.04,0.04,-0.015,-0.015,0.30,0.005,0.005,0.015\rbrace.\nonumber
\end{eqnarray}
Since $\bar{\omega}_{6}$, $\bar{\omega}_{15}$ respectively, reflects the
correlation of the MM signal with its corresponding PM. As expected its
magnitude is greatest among all parameter values. Next, the MM intensity
shows strong correlations with its direct nearest neighbors positioned at
$(x\pm1,y)$,~i.e. $\bar{\omega}_{2/11}$ and $\bar{\omega}_{3/12}$. Hence,
the stronger the MM-neighboring intensities, the stronger their influence
on the background signal of MM.  However, in order to somehow compensate
for strong MM-neighboring signals which might be caused by the presence
of complementary target sequences, parameters $\bar{\omega}_{4/13}$
and $\bar{\omega}_{5/14}$ have negative sign. The remaining three
parameters,~i.e.~the top three neighbors limit their influence to a
basically negligible minimum which appears to indicate the weak sequence
correlation in $y$-direction as previously found.  The results indicate
that the influence of neighboring intensities on the background noise
is significant. In fact, our analysis shows that $\approx 30\%$ of
$\log I_{i,bg}^{est}$ are constituted by neighboring probes in terms of
absolute intensities (see Table~\ref{Tab:local_percentage}). It appears that for a few probes,
the latter might even play a crucial role.

It is unlikely that the correlations summarized in Fig~\ref{fig:neighbors_350}
could be explained only as simple optical effects,
as light from bright probes spilling into their neighborhood. 
Indeed, optical effects
should produce an isotropic correlation pattern in the $x$ and $y$
directions which is not seen in our analysis. Another cause for this
correlation might be that neighboring probes share common sequences,
as is the case in Affymetrix chips~\cite{krol08}.

\begin{table}[h!]
\centering
\begin{tabular}{l|ll}
X. Laevis & 74500.CEL & 76190.CEL \\ 
 & 35\% & 37\% \\ 
\hline
C. Celegans & 201989.CEL & 201994.CEL \\ 
 & 43\% & 52\% \\ 
\hline
H. Sapiens & 263931.CEL & 263930.CEL \\ 
 & 41\% & 39\% \\ 
\hline
S. Cerevisiae & 207569.CEL & 207570.CEL \\ 
 & 29\% & 29\%
\end{tabular}{} 
\caption{Influence of neighboring spot on background intensity in \%
of eight (randomly chosen) CEL-files of different organisms.}
\label{Tab:local_percentage}
\end{table}


\subsection*{Nearest-Neighbor Parameters}

The parameters $\bar{\omega}_{\alpha}$ ($19\leq\alpha\leq 34$) are
the analogues of the nearest-neighbor free energy parameters. The
nearest-neighbor model is commonly used to study the thermodynamics
of hybridization of nucleic acids in solution (see e.g. \cite{bloo00}). In
this model it is assumed that the stability and thus the hybridization
free energy $\Delta G$ of a dinucleotide depends on the orientation and
identity of the neighboring base pairs. For RNA/DNA duplexes there are
16 hybridization free energy parameters which were measured in aqueous
solution by Sugimoto et al.~\cite{sugi95}.

Recent experiments~\cite{hooy09} focusing on specific hybridization show
a good degree of correlation between the hybridization free energies in
solution and those directly determined from microarray data. Concerning
background data, we also expect a certain degree of correlation between
the parameters $\bar{\omega}_{\alpha}$ ($19\leq\alpha\leq 34$) and 
their corresponding Sugimoto free energy parameters.

In order to test the relationship between the experimentally determined
so-called Sugimoto parameters and our results, we calculate the
correlation coefficient between these two sets. The results are reported
in Table~2. In general, the correlation coefficients vary between $0.53$
and $0.83$ with a median value of $0.71$. Figure~2 shows two typical
results for C. Elegans and D. Melanogaster. Both plots indicate that
the relationship between $\bar{\omega}_{\alpha}$ ($19\leq\alpha\leq
34$) and the nearest-neighbor free energy parameters of~\cite{sugi95}
is approximately linear.

\begin{figure}[!t] 
\centerline{\includegraphics[width=12cm,height=16cm]{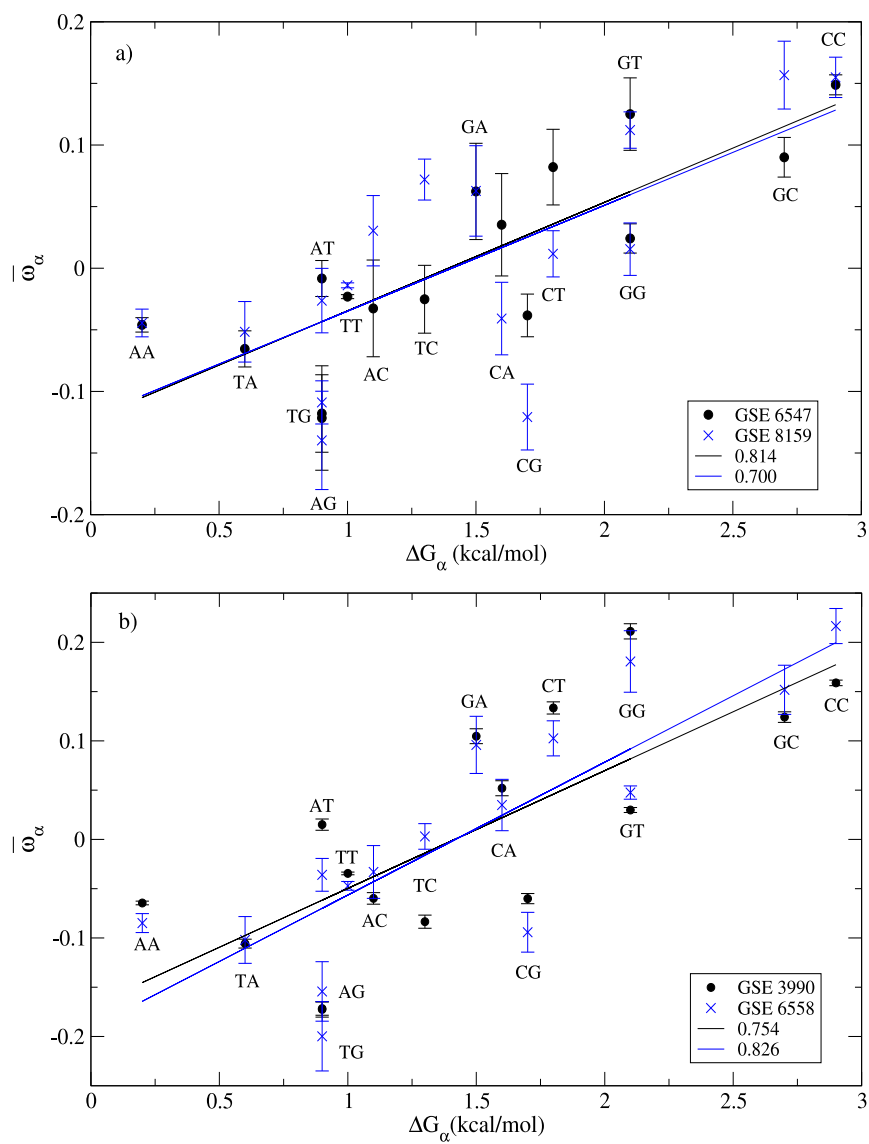}}
\caption{\textbf{Fitted vs. solution hybridization free energies:}
Pair Parameters $\bar{\omega}_{\alpha}$ ($19\leq\alpha\leq 34$),
as obtained from the minimization of Eq.~(\ref{eq:lin_eq}) as function of
$\Delta G_{\alpha}$, the nearest-neighbor stacking free energy obtained
from DNA/RNA hybridization in solution~\cite{sugi95}. Plots refer
to two GSE training data sets of a) C. Elegans and b) Drosophila
Melanogaster. Each point on the $y$-axis is the averaged value
over all CEL-files representing its GSE-set. The error bars are the
standard deviation.  Notation of DNA pairs are from 5' to 3' end. The
straight lines are linear fits; correlation coefficients of each GSE
set in legend.}
\label{fig:sugi_worm_fly} 
\end{figure} 

\subsection*{Position-dependent nearest neighbor parameter}

Figure~\ref{fig:parabsugi_worm_fly} shows a plot of position-dependent nearest-neighbor parameters
$\bar{\omega}_{35}$ to $\bar{\omega}_{50}$ as obtained from the
minimization of the cost function of Eq.~(\ref{eq:function_s}) for four
different sets of experiments. The data are plotted as function of the
corresponding nearest neighbor free energy parameters of~\cite{sugi95}. We
note that in all experiments shown there is a negative correlation between
the two data sets. The parameters $\bar{\omega}_{35}$ to $\bar{\omega}_{50}$
reflect the difference in effective free energy between the ends and the
middle of the probe sequence. Since weakly binding probes suffer more from
end-effects (unzipping, etc.), $\bar{\omega}_{35}$ to $\bar{\omega}_{50}$ 
and their corresponding nearest-neighbor free energy parameters are 
negatively correlated. Hence, the negative correlation in Fig.~3
and those observed in all other cases (see correlation coefficients
reported in Table~2) indicates that the dominant contribution to the
background intensity comes from the middle nucleotides. This conclusion
complies with other types of analysis which use position-dependent
free energy parameters (see e.g.~\cite{bind05,zhan03}).

\begin{figure}[!t] 
\centerline{\includegraphics[width=12cm,height=16cm]{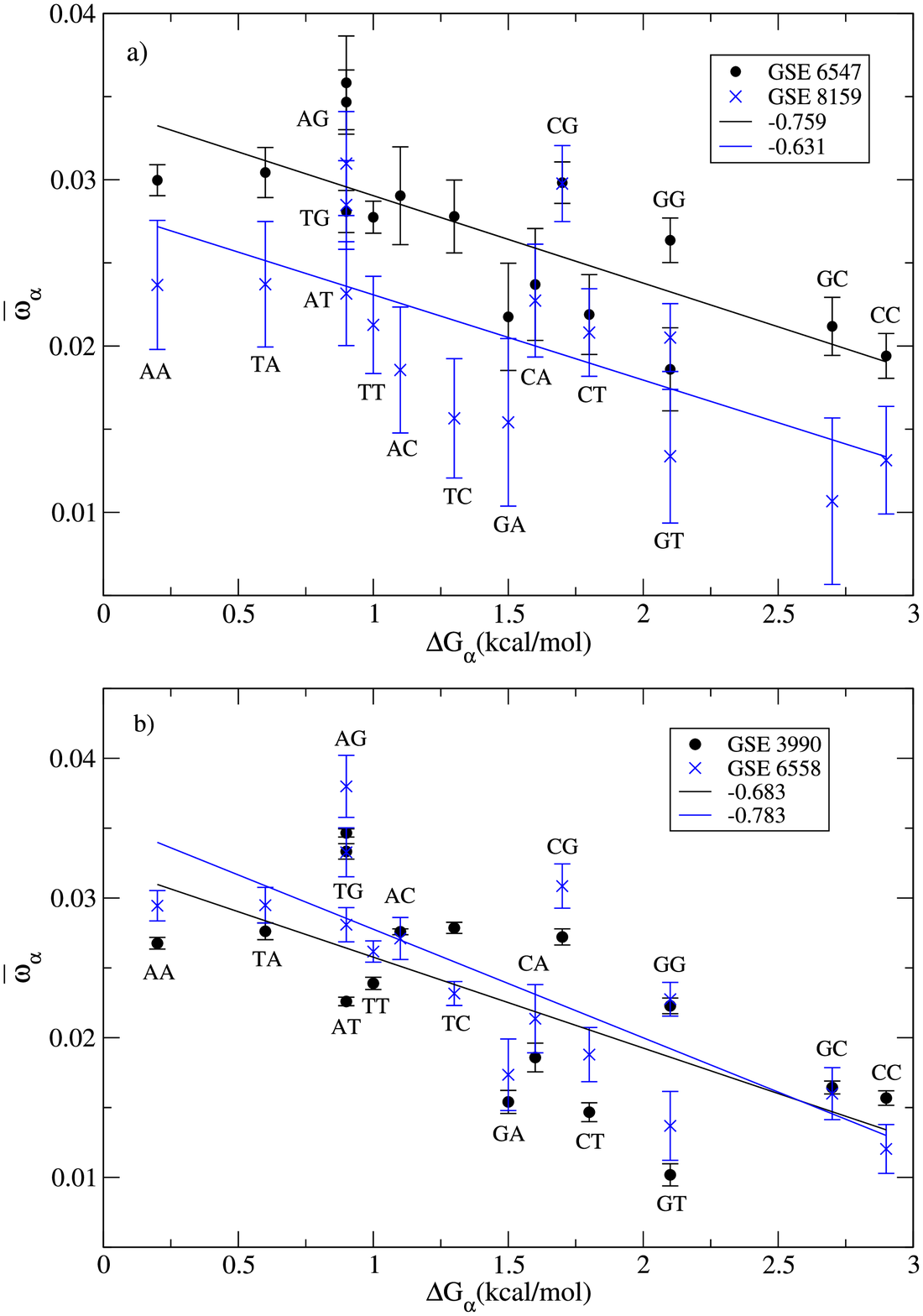}}
\caption{\textbf{Fitted vs. solution hybridization (position
dependent) free energies: }Same as in Fig.~\ref{fig:sugi_worm_fly} for parabolic pair
parameters,~i.e.  $\Delta G_{\alpha}$ vs. $\bar{\omega}_{\alpha}$
($35\leq\alpha\leq 50$) for a) C. Elegans and b) D. Melanogaster}
\label{fig:parabsugi_worm_fly} 
\end{figure} 

\subsection*{Comparing estimated vs. measured background}

In this section we compare the estimated background signal (as given in
Eq.~(\ref{eq:back_est})) with the intensities of given probe
sets corresponding to non-expressed genes in the samples analyzed.
We start with publicly available data taken from spike-in experiments
on HGU95A chips (www.affymetrix.com) where genes have been spiked-in at
known concentrations, ranging from $0$ to $1024$ pM (picoMolar). The
data at $0$ pM correspond to the absence of transcript in solution.
Figure~\ref{fig:37777and209606}a compares the measured and predicted background for probeset
37777\_at. Except for probes 2 and 15 for which the measured signal is
higher than the predicted background, there is a nice agreement between
our prediction and the experimental background intensity: the standard
deviation of the absolute difference between the intensity of the PM
and $I^{\rm est}$ is $28$ in Affymetrix intensity units. Also shown are
the background estimations as obtained with other algorithms (the data
for PDNN is missing in Fig.~\ref{fig:37777and209606}a due to the unavailability of one of the
supporting files from Bioconductor packages). Figure~\ref{fig:37777and209606}b presents the
same information for probeset 209606\_at. Here, the standard deviation
is $7$ in Affymetrix intensity units.
\begin{figure}[!t] 
\centerline{\includegraphics[width=12cm,height=16cm]{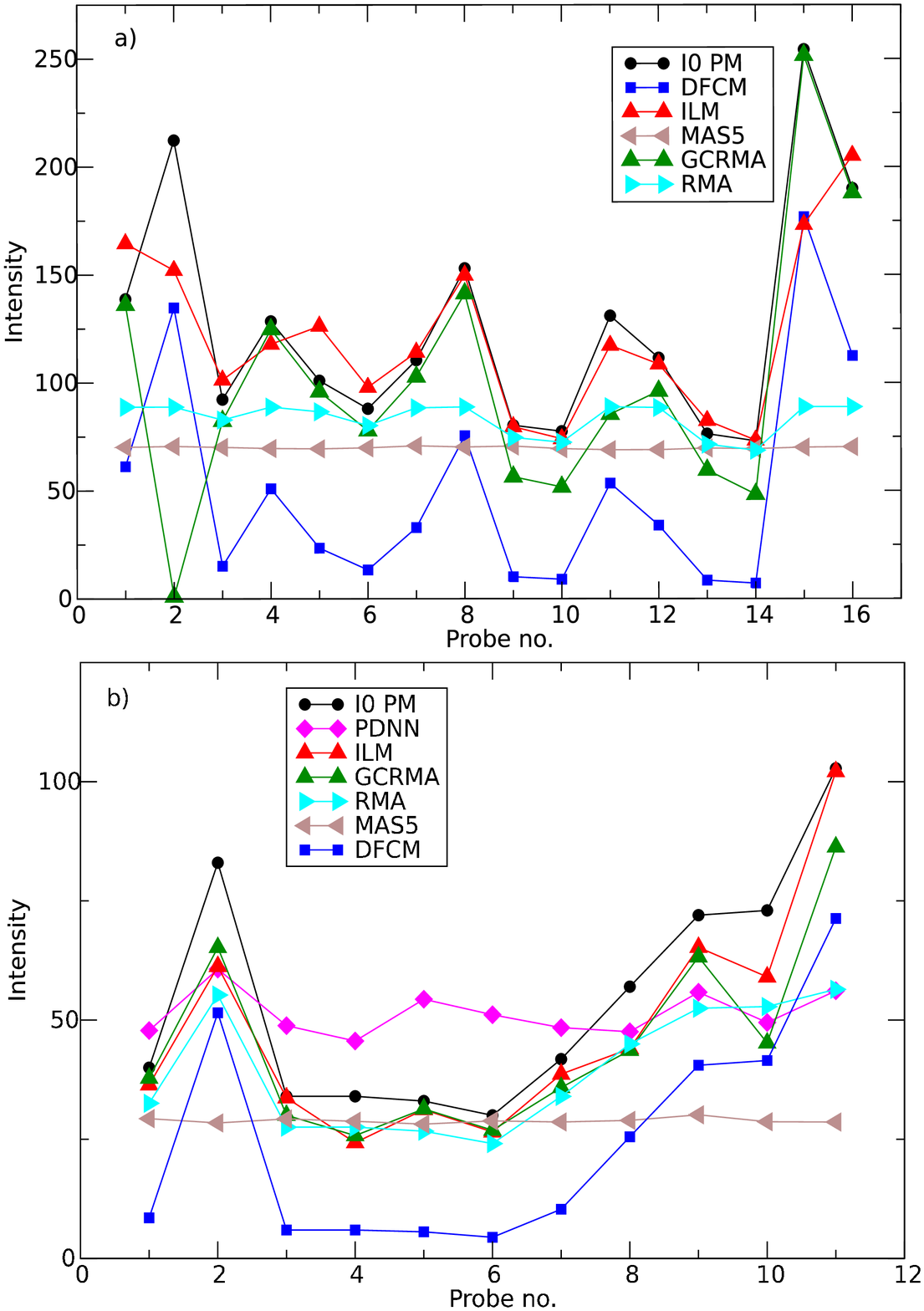}}
\caption{\textbf{Computed vs. true background (spike-in
experiments)}Experimental results ($I0PM$) and theoretical
predictions of the algorithms discussed here for a) probeset
37777\_at (HGU95A, 1521a99hpp\_av06.CEL) and b) 209606\_at
(HGU133A,12\_13\_02\_U133A\_Mer\_Latin\_Square\_Expt10\_R1.CEL) spiked-in
at concentration $c=0$, i.e. absent from the hybridizing solution.
All comparison were performed using freely available packages from the
Bioconductor project.}
\label{fig:37777and209606} 
\end{figure} 

Next, we go beyond the spike-in experiments. To select non-expressed
genes we considered probe sets with very low expression values as
obtained from the RMA algorithm. In Figure~\ref{fig:AThandCEl} the PM intensities as well
as the calculated background signal of two probesets from A. Thaliana
and C. Elegans experiments are shown. The absolute intensities of
both probesets are very low. As a consequence, we can safely 
assume these genes are not expressed and hence any measured signal 
can be attributed to the background. 

\begin{figure}[!t] 
\centerline{\includegraphics[width=12cm,height=16cm]{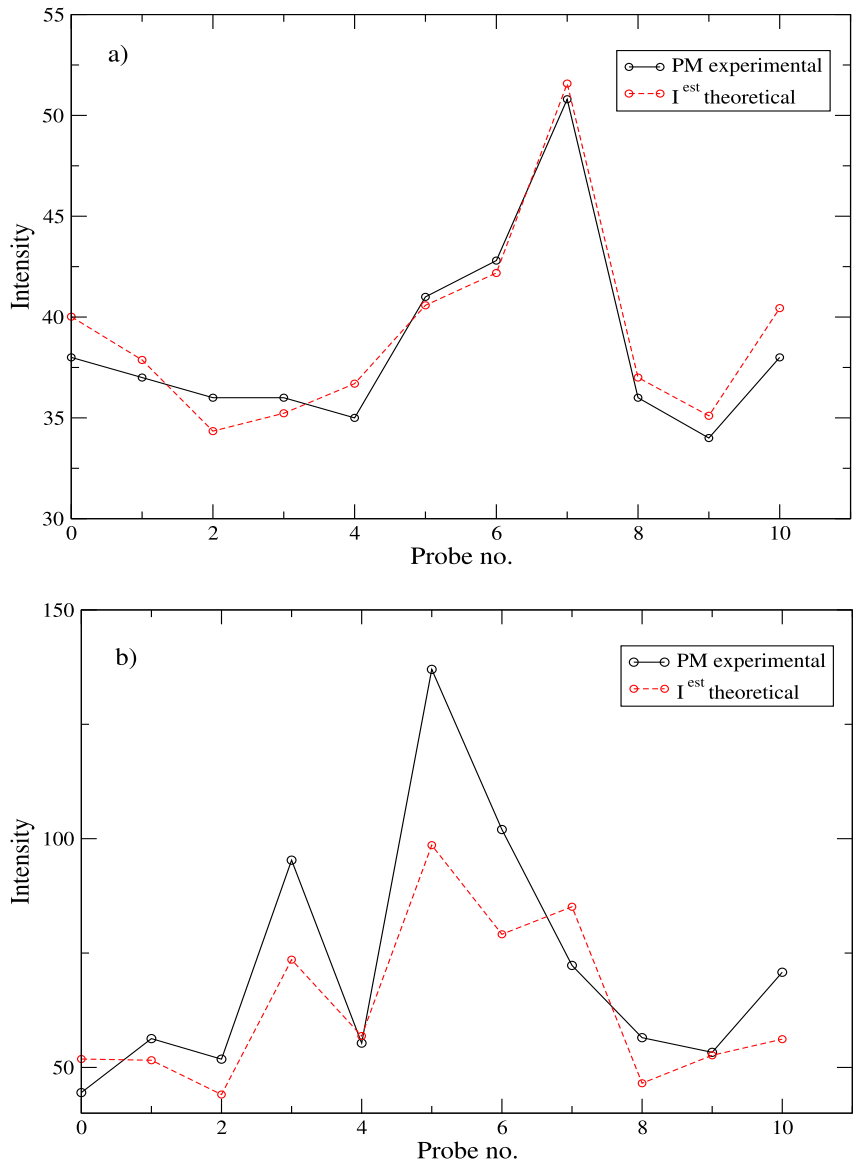}}
\caption{\textbf{Computed background vs. low expressed genes data:} Experimental results (solid black line) and theoretical prediction
(dashed red line) of probeset a) 256610\_at (A. Thaliana, GSE4847,
GSM109107.CEL) and b) 175270\_at (C. Elegans, GSE8159, GSM201995.CEL)}
\label{fig:AThandCEl} 
\end{figure} 
Figures~\ref{fig:37777and209606} and~\ref{fig:AThandCEl}  both show that 
the present model captures the essentials of and correctly predicts 
background intensities. Both Figures are representative for the 
CEL-files analyzed in this work.

\section*{Discussion}
We have presented a background subtraction scheme for Affymetrix
GeneExpression arrays which is both, accurate and usable on a standard
x86\_64 Intel Core 2 PC. The algorithm centers around a cost function
which is quadratic in its fitting parameters. This allows for a rapid
minimization, through linear algebra, in particular through singular
value decomposition. The accuracy of the present algorithm is very
similar to that of a background algorithm previously presented by the authors
\cite{krol08}. The latter had been tested on Affymetrix spike-in data and its 
performance was compared to background schemes such as MAS5, RMA and 
GCRMA. Regarding spike-in data, the analysis had shown that the 
proposed algorithm is definitely more accurate than background 
computations done with MAS5 and RMA, but also improves on GCRMA~\cite{krol08}.

The proposed algorithm has two categories of fitting parameters. The
first category exploits correlations between features which are neighbors
on the chip. The second category is based on the strong similarity
between probe-target hybridization and duplex stability in solution, and
involves stacking free energies in analogy to those in the nearest-neighbor
model. Existing algorithms are either of the first ~\cite{affy01}
or the second~\cite{wu04,iriz06,zhan03} category, but not both.

The background subtraction scheme has been tested on 360 GeneChips
from publicly available data of recent expression experiments. Since
the fitted values for the same parameters in different experiments
do not show much variation, the algorithm is robust and can be easily
transferred to other experiments.

Due to its speed and accuracy the present method is suited for
large scale computations. An R-package integrating the background
analysis scheme with the computation of expression values from
background subtracted data will be made freely available to the
community (a preliminary version of this package can be found in
\url{http://itf.fys.kuleuven.ac.be/~enrico/ilm.html}).  The performance
of this approach is discussed in \cite{muld09}.

\section*{Competing interests}
The authors declare that they have no competing interests.

\section*{Authors contributions}
EC and GB planned and supervised the research. KMK wrote the code for
data analysis and analyzed the data. KMK and EC wrote the paper.

\section*{Acknowledgements}
  \ifthenelse{\boolean{publ}}{\small}{}
We acknowledge financial support from FWO (Research Foundation - Flanders)
grant n. G.0311.08.  Stimulating discussions with N. Naouar are gratefully
acknowledged.




\end{bmcformat}
\end{document}